\documentclass[sigconf]{acmart}
\copyrightyear{2022}
\acmYear{2022}
\setcopyright{acmcopyright}
\acmConference[SIGIR '22]{Proceedings of the 45th International ACM SIGIR Conference on Research and Development in Information Retrieval}{July 11--15, 2022}{Madrid, Spain.}
\acmBooktitle{Proceedings of the 45th International ACM SIGIR Conference on Research and Development in Information Retrieval (SIGIR '22), July 11--15, 2022, Madrid, Spain}
\acmPrice{15.00}
\acmISBN{978-1-4503-8732-3/22/07}
\acmDOI{10.1145/3477495.3531660}
\AtBeginDocument{%
  \providecommand\BibTeX{{%
    \normalfont B\kern-0.5em{\scshape i\kern-0.25em b}\kern-0.8em\TeX}}}
\usepackage[utf8]{inputenc}
\usepackage{graphicx}
\usepackage{graphics}

\begin{document}
\fancyhead{}

\title{DIANES: A DEI Audit Toolkit for News Sources}

\author{Xiaoxiao Shang{$^1$}, Zhiyuan Peng{$^1$}, Qiming Yuan{$^1$}, Sabiq Khan{$^1$}, Lauren Xie{$^1$}, Yi Fang{$^1$}}
\authornote{Corresponding authors}
\author{Subramaniam Vincent{$^2$}}
\authornotemark[1]
\affiliation{
%\institution{Santa Clara University}
\institution{{$^1$}Department of Computer Science and Engineering}
\institution{{$^2$}Markkula Center for Applied Ethics}
\city{Santa Clara University}
\state{California}
\country{USA}}
\email{{xshang,zpeng, qyuan2, skhan2, lxie, yfang, svincent}@scu.edu}

\begin{abstract}
Professional news media organizations have always touted the importance that they give to multiple perspectives. However, in practice, the traditional approach to all-sides has favored people in the dominant culture. Hence it has come under ethical critique under the new norms of diversity, equity, and inclusion (DEI). When DEI is applied to journalism, it goes beyond conventional notions of impartiality and bias and instead democratizes the journalistic practice of sourcing -- who is quoted or interviewed, who is not, how often, from which demographic group, gender, and so forth. There is currently no real-time or on-demand tool in the hands of reporters to analyze the persons they quote. In this paper, we present DIANES, a DEI Audit Toolkit for News Sources. It consists of a natural language processing pipeline on the backend to extract quotes, speakers, titles, and organizations from news articles in real time. On the frontend, DIANES offers the WordPress plugins, a Web monitor, and a DEI annotation API service, to help news media monitor their own quoting patterns and push themselves towards DEI norms.
\end{abstract}




\maketitle

\section{Introduction}

Diversity, equity, and inclusion (DEI) are fundamental to promoting robust journalism that supports a healthy society, by fostering well-researched, complex stories that explore different perspectives and voices. As the world is becoming more diverse, it is the news media's responsibility to reflect this. Consequently, sourcing in news is crucial since the sources that journalists choose to quote in their stories affect whose stories get told, how stories are told, whom the news is for, and what communities are served. Some studies show that the voices of women and minorities are often substantially under-quoted in media stories \cite{asr2021gender}. Thus, it is important for newsrooms to track who is quoted, how often they are quoted, what are the proportions of people quoted by gender, title, race, ethnicity, community, etc. However, for everyday reporters and editors who are part of large and small newsrooms, there is no daily system to monitor their own quoting patterns and push themselves towards the DEI norms. On the other hand, the recent advances in information retrieval and natural language processing have enabled deeper understanding of document content, which opens opportunities for (semi)-automatic DEI auditing based on computational approaches. 

In this paper, we present DIANES, a DIversity Auditor for News Sources. DIANES is a DEI toolkit consisting of an NLP pipeline on the backend to extract quotes, speakers, titles, and organizations from news articles in real time. On the frontend, DIANES supports a variety of user requests on the DEI audit based on the information extracted on the backend by offering the WordPress plugins, a Web monitor, and an annotation API service. The toolkit can help reporters visualize source-diversity proportions (e.g., gender and race) for quotes in their article drafts as well as published pieces. Moreover, DIANES can inform editors of the quoting patterns of all the published articles by their newsrooms or across multiple sites. In addition, the annotation API can extract the relevant information from any news articles for downstream applications that newsrooms or other parties may wish to develop by themselves to support their specific DEI goals. Unlike one-time, manual audits, DIANES provides nearly on-demand feedback to ease barriers to assessing stories’ representativeness and it may offer immediate opportunities to fix inequitable reporting. DIANES is currently in test use by several newsrooms. To the best of our knowledge, DIANES is the first toolkit for auditing DEI of news sourcing including gender and race/ethnicity.

\begin{figure*}[t]
    \centering
    \includegraphics[width=\linewidth]{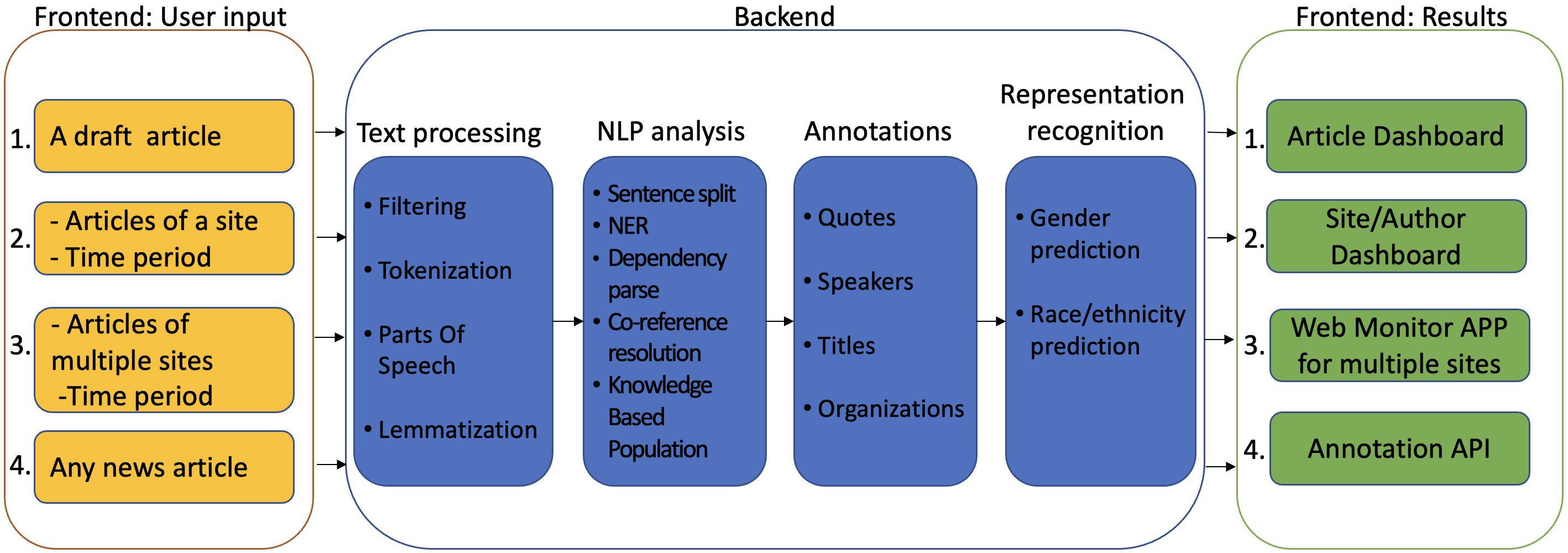}
    \caption{The diagram with the major components of DIANES.}
    \label{fig:flow}
\end{figure*}

\section{Related Work\label{sec:rw}}
Sources bring credibility and authority to news reports \cite{reich2010source}. Traditional news sourcing practices that favor official and dominant voices have been widely documented over time and across media. The Global Media Monitoring Project in 2010 \cite{macharia2010makes} studied nearly 1,300 newspapers, television and radio stations in 108 countries and found fewer than one in four news subjects were women. Results from the Project for Excellence in Journalism in 2005 \cite{artwick2014news} obtained similar results. The earlier findings were even more extreme, showing that only 10 percent of sources in newspaper front-page stories were women \cite{brown1987invisible}.

Some computational systems have been developed for newsrooms to track the demographics of their sources. With a tool named Dex \cite{DEX2021}, reporters and editors in National Public Radio (NPR) can submit information about their sources and later pull up reports to monitor their source diversity. The manual approaches are often not scalable. To examine gender bias in media, the Gender Gap Tracker \cite{asr2021gender} was recently created to tally the number of men and women quoted in news text using natural language processing. However, this system does not support real-time processing of news articles nor cover the important dimension of race and ethnicity in diversity.

\section{System Design\label{sec:sys}}
DIANES supports a variety of user requests which are processed by a natural language processing pipeline on the backend and then results are returned to the frontend. The diagram with the major components of DIANES is shown in Figure \ref{fig:flow}.

\subsection{Frontend}
DIANES currently accepts four different types of user requests below: 

\begin{enumerate}
\item When writing an unpublished draft article on WordPress, a reporter can request to see all the quotes, speakers, their titles, organizations, gender and race extracted from the draft.
\item Given all the news articles published by a newsroom in a time period, editors can request to see the gender and race distributions of the news sources (quotes) for a given author or the whole newsroom.
\item Users can request to see the gender and race distributions across multiple newsrooms/sites.
\item DEI application developers can access the relevant DEI information extracted from any given article to develop their own interfaces or downstream applications.
\end{enumerate}

For the above requests 1 and 4, users input a single news article, which is wrapped into a JSON file along with an article ID and authentication key. For the requests 2 and 3, user inputs are news archives (often in the XML format) given by different news sites or data providers. We extract relevant information from the archives and store it in a database for subsequent processing. 

Given a user request/input, DIANES will return and show the respective results as described below:

\subsubsection{Article Dashboard.} DIANES provides a WordPress plugin, which works in conjunction with a news article annotation server to identify quotes in the text of stories in real time. After the reporter saves the draft article, they can request the DEI data from the backend. Then we insert all the information in the JSON result returned by the backend into the corresponding tables that are created in the WordPress database. Next we show DEI tables of quotes for reporters to review. DIANES calculates the gender and titled person proportions and uses Google Charts\footnote{\url{https://developers.google.com/chart}} to display these proportions.

\subsubsection{Site/Author Dashboard.} 
Site/Author Dashboard is also included in the WordPress plugin. Specifically, we process archives from newsrooms in bulk for bootstrapping DEI data. When data is ready from the backend, we will insert them into the WordPress database. We can then view the relevant data (such as the proportions of genders, races, titled speakers, etc.) in the most recent month by default or in a different month that the user selects. After loading the data into variables, we use Google Charts to display the results on the frontend.

\subsubsection{Web Monitor for Multiple Sites.}
In addition to the WordPress plugin, we have prototyped a web monitor application that offers the same visualizations at the site level for thousands of news sites in the U.S. The web monitor is an application hosted and supported by React, Node.js, Express, and MySQL. We use Node.js and Express as the backend server to connect with the MySQL database, and use React.js for the user interface. This monitor application allows user to access the DEI data visualization for authorized news sites.

\subsubsection{Annotation API} 
The Annotation API is an endpoint for news article analysis through HTTP requests. The API consists a simple Flask web server. The input is a JSON file that contains the article to be analyzed. The web API will return the results in JSON as well. All the quotes and their speakers' names, genders, titles, and community representations are included in the results.

\subsection{Backend}
The backend module takes news article(s) as input and annotates the quotes and speakers with their titles and organizations, and then predicts gender and race of the speakers. It consists of the following four stages and the first two stages leverage the Stanford CoreNLP library\footnote{\url{https://stanfordnlp.github.io/CoreNLP/}} \cite{manning-etal-2014-stanford}.  

\subsubsection{Text Processing.} 
%Also, timeouts can happen for very long paragraphs. To avoid this, paragraphs with more than 500 words will be removed.

We first filter some extraneous information from the news articles such as XML tags. After the filtering, the document will be tokenized. 
The next step is the part of speech (POS) tagging which assigns POS labels to tokens, such as whether they are verbs or nouns, and then lemmatization is applied to map a word to its dictionary form.

\subsubsection{NLP Analysis.\label{sec:nlp}} DIANES conducts sophisticated NLP analysis on every article through the CoreNLP server. It contains the following specific components.

Sentence splitting is the process of dividing text into sentences. CoreNLP splits article text into sentences via a set of rules \cite{manning-etal-2014-stanford}. Named Entity Recognition (NER) annotator is used to extract person names and organizations by using machine learning sequence models. To extract titles, we added the title pattern file in CoreNLP so that the NER annotator could recognize the titles based on a set of rules.

Dependency parsing analyzes grammatical relations between words in a sentence and extracts textual relations based on the dependencies which are triplets: name of the relation, governor, and dependent \cite{nivre2016universal,schuster2016enhanced}. The co-reference resolution finds mentions of the same entity in a text, such as when ``Anne'' and ``she'' refer to the same person. Knowledge Base Population (KBP) annotator \cite{angeli2015bootstrapped, zhang2016stanford} extracts relation triples meeting the TAC-KBP\footnote{\url{https://tac.nist.gov/2017/KBP/}} specifications. We used it to find titles and organizations of a person if presented in the text. Dependency parsing and co-reference resolution are required to extract KBP relations and quotes in the NLP pipeline.

\subsubsection{Annotations.} In this stage, we produce annotations for quotes, speakers, and their titles and organizations based on the NLP analysis. If a quote has a missing quotation mark, CoreNLP would continue searching for the closing quotation mark and include everything in-between as the quote's content, which may decrease quote resolution accuracy. We address such a situation by looking at the number of words in the quotes. Quotes are considered as long quotes if they contain more than 100 words. On the other hand, quotes with less than 5 words are considered as short quotes and will be dropped since those quotes may increase the mis-resolution rate of speakers and such cases may come from book titles or slogans containing irrelevant information. 

For each person in the article, CoreNLP may detect different titles for him or her based on the contexts, where we only record the first title detected for that person. If the speaker of the quotes shares the same last name with other persons, CoreNLP would provide wrong co-reference some times. In such cases, we re-annotate the text from the beginning of the article till the paragraph where the quote is, to ensure the accuracy of speaker detection.
We also find that CoreNLP has a high probability of mistaking the speaker based on their approach to processing quotes \cite{muzny2017two} if the quotes' attribute had some particular patterns. When these patterns occur, we mark the speaker as doubtful in the results.

\subsubsection{Representation Recognition.} In this stage, we predict the representations of the speakers based on the annotations extracted from the previous stage. DIANES currently supports two important representation attributes of DEI: gender and race/ethnicity, as they are the crucial information for understanding the representativeness of news sourcing.

Approaches to gender tagging in the text have majorly been database-reliant \cite{santamaria2018comparison}. The key idea behind these approaches is to maintain a database of first names against which occurrences of named entities are compared \cite{das2021context}. We use Gender API\footnote{\url{https://gender-api.com}} as our name-to-gender inference service since this service demonstrated competitive results in the benchmark evaluations \cite{santamaria2018comparison}. More details can be found in \cite{santamaria2018comparison} which also introduced other popular gender inference services.

There is much less existing work on predicting the race/ethnicity of a person given a name, and the task is more challenging than gender detection. Thus, we implemented and trained a machine learning model for this task. Our model accepts the name as input, and outputs the probabilities of the predictions which are converted to a confidence score for the predictions. A small confidence score can alert the end users about the potentially inaccurate results. Specifically, our model encodes the name as a sequence of bi-grams, passes them through Bidirectional LSTM \cite{schuster1997bidirectional}, connect it with a dense layer, and the softmax layer to produce the probabilities. The model is similar with \cite{sood2018predicting}.

The dataset we used to train our race detector was extracted from the United States Census Bureau in 2000 and 2010\footnote{\url{https://www.census.gov/topics/population/genealogy/data/2010_surnames.html};
\url{https://www.census.gov/topics/population/genealogy/data/2000_surnames.html}}, which contains a total of 151,670 unique names in 6 categories (White, African American, American Indian and Alaska Native, Asian, and Native Hawaiian and Other Pacific Islander). The 2-gram based vocabulary yielded 962 different 2-grams. We trained two models with one for binary classification (White vs. non-White) and another one for six-category classification. 

% Once the representation recognition step is done, the backend generates a resulting JSON object that contains all the quotes, speakers, titles, organizations, and annotations, and sends the object back to the frontend client.

 \begin{figure*}[ht]
    \centering
    \includegraphics[width=\linewidth]{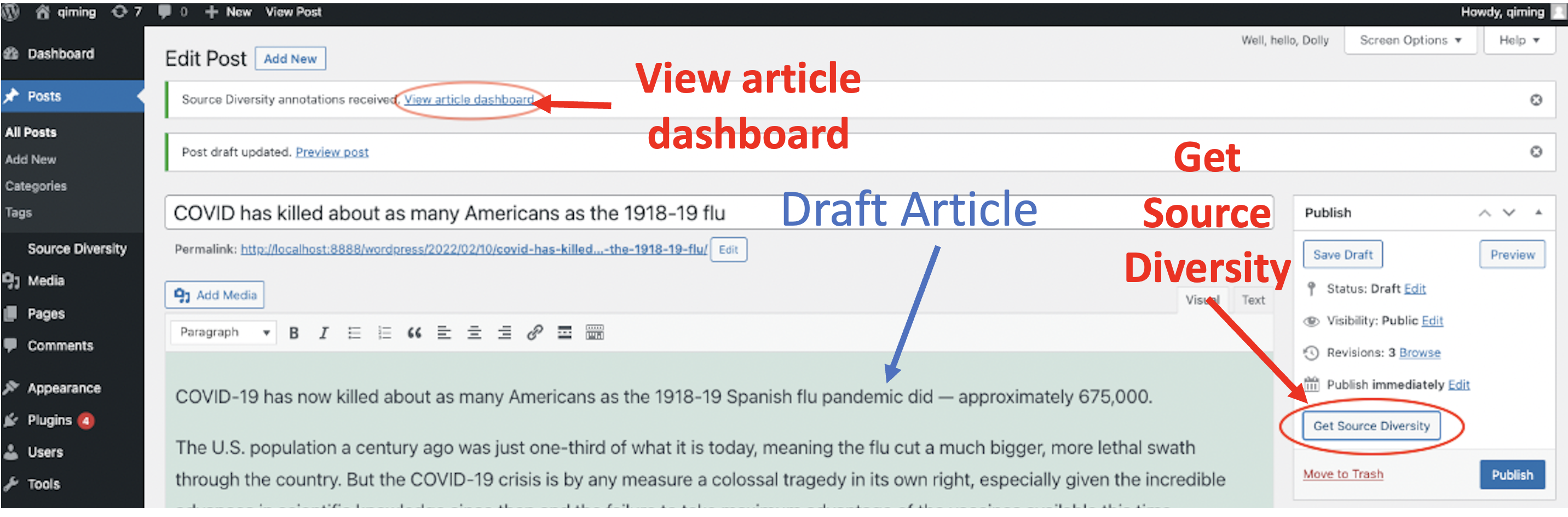}
    \caption{Through the WordPress plugin, a reporter can send requests to get the source diversity information when writing a draft news article.}
    \label{fig:UI}
\end{figure*}

\begin{figure*}[ht]
  \centering
  \includegraphics[width=\linewidth]{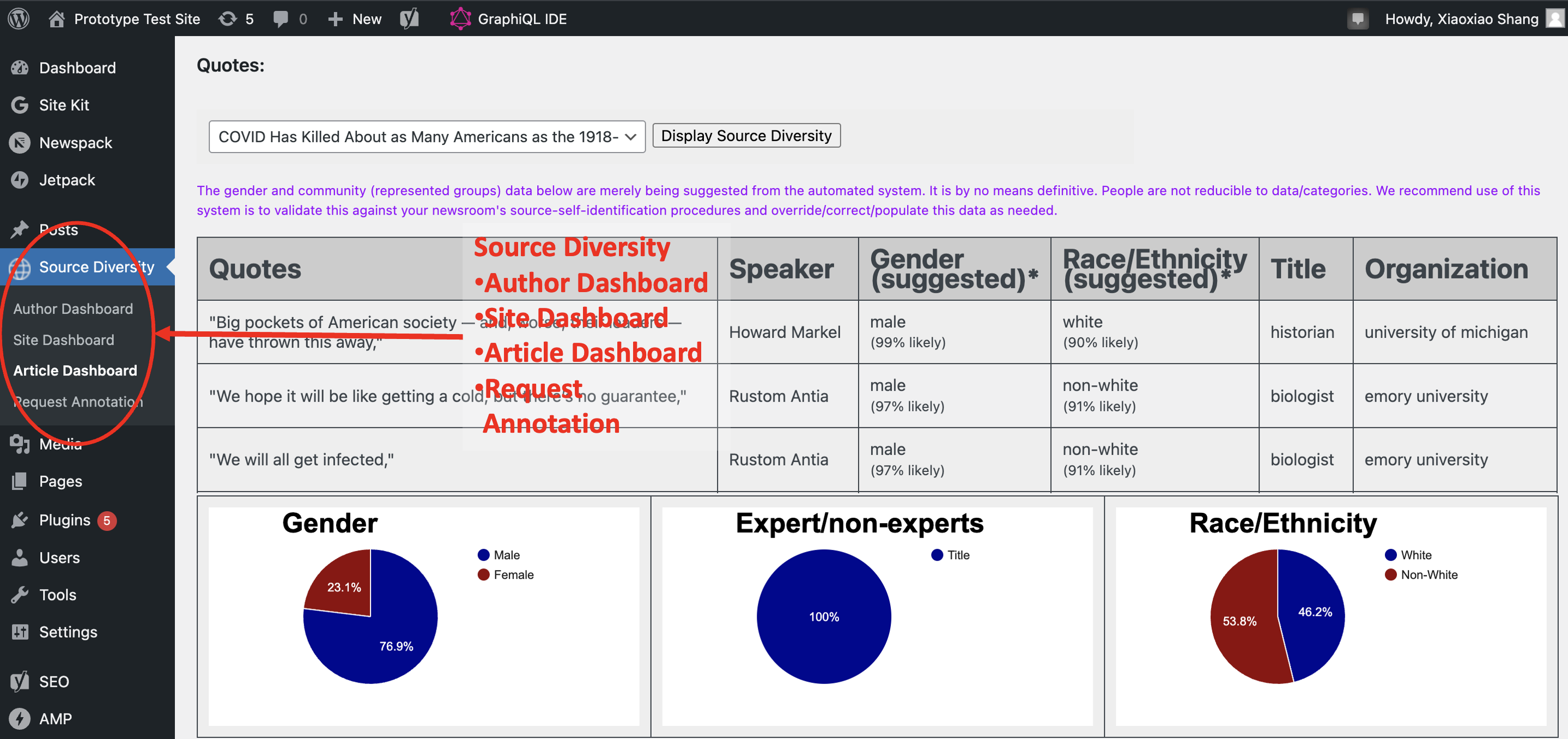}
  \caption{The Article Dashboard shows a table of quotes displayed with DEI data after a reporter or an editor requests annotation of a draft or a published article.}
  \label{fig:dei}
\end{figure*}

\subsection{Evaluation}

To evaluate the accuracy of the proposed race prediction model, we split the data of 2 million names collected from the US Census into 72\% for training, 8\% for development, and 20\% for test, which resulted in 160,000 names in the test set. The accuracy for the binary classification (White vs non-White) is 82\% and the accuracy for the 6-category classification is 81\%.

We also created two datasets to evaluate other backend modules. One is the Mainline news sources with the topics about homelessness, which was collected at San Francisco State University. We randomly chose 20 articles from it. Another dataset is 30 articles randomly sampled from FakeNewsCorpus\footnote{\url{https://github.com/several27/FakeNewsCorpus/releases/tag/v1.0}}. We manually labeled these news articles. The numbers of name occurrences in these two datasets are 375 on Mainline and 86 on FakeNewsCorpus, respetively. The numbers of quotes are 409 on Mainline and 95 on FakeNewsCorpus, respectively. The accuracy for the speaker resolution component was 92\% on Mainline and 86\% on the FakeNewsCorpus source. The accuracy for the title extraction component was 80\% on Mainline and 67\% on the FakeNewsCorpus source. The accuracy for the Gender API was 92\% on the European name corpora based on the published results in the literature \cite{santamaria2018comparison}.

\section{Demonstration}
In this section we briefly demonstrate the main functionalities of DIANES. A short video recording of the demonstration can be found here\footnote{\url{https://www.youtube.com/watch?v=RC5ielXO3Wo}}. 

\subsection{Article Dashboard}
While a reporter is writing a draft article on WordPress, the reporter can request to get the DEI data from DIANES' backend server by clicking on ``Get Source Diversity'' as shown in Figure \ref{fig:UI}. This is called on-demand article annotation. As soon as the backend server completes the processing, the ``View article dashboard'' button will be enabled. After clicking on the button, a table of results will be shown for each captured quote and its speaker's name, title, organization, and community representation, as demonstrated in Figure \ref{fig:dei}, with pie charts generated to show the distributions of the DEI attributes of interest.

\begin{figure}[t]
  \centering
  \includegraphics[width=\linewidth]{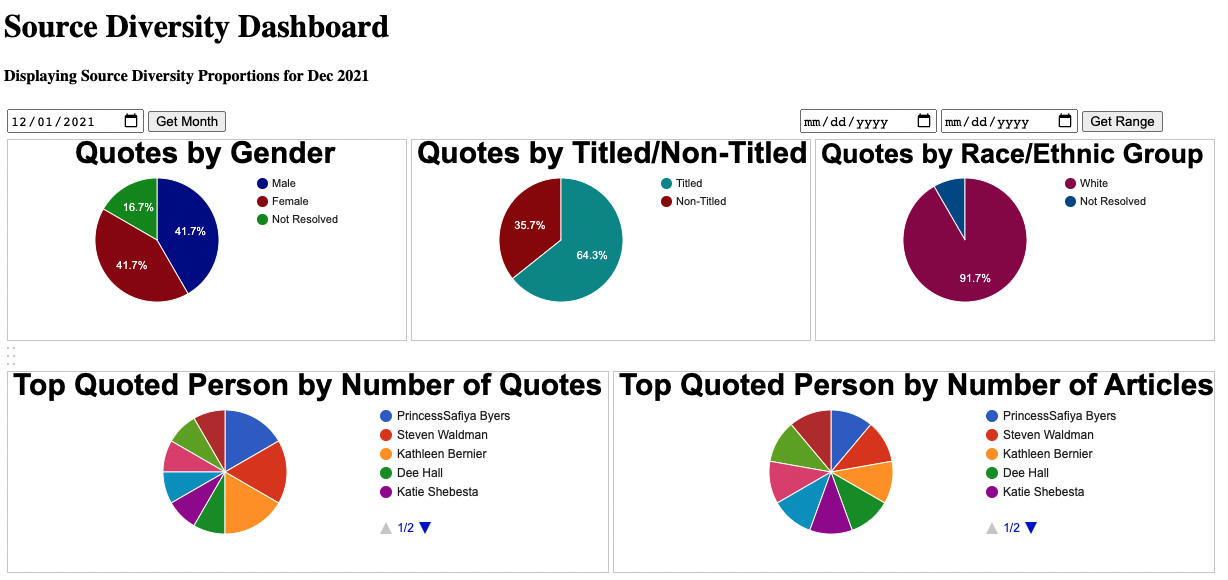}
  \caption{The Site/Author Dashboard displays the DEI annotated data for all the articles published in a specified time period from the news site or from a given author. The display also includes top-quoted persons.}
  \label{fig:monthproportion}
\end{figure}

\subsection{Site and Author Dashboards\label{sec:dash}}
If a user clicks on ``Site Dashboard'' in Figure \ref{fig:dei}, it will bring the user to the source diversity dashboard for the newsroom as shown in Figure \ref{fig:monthproportion}. Users such as editors can select a month or a time range and get the visualization of the DEI data for all the articles published in that time period by the newsroom. It loads in the most recent month's data by default. Similarly, an author can retrieve the DEI data for the articles he or she wrote in a given time period by using the Author Dashboard. The result visualization is very similar to Figure \ref{fig:monthproportion} while it is generated based on the author instead of the whole site. This will give authors an intuitive understanding of their own quoting patterns.

\begin{figure}[t]
 \centering
 \includegraphics[width=\linewidth]{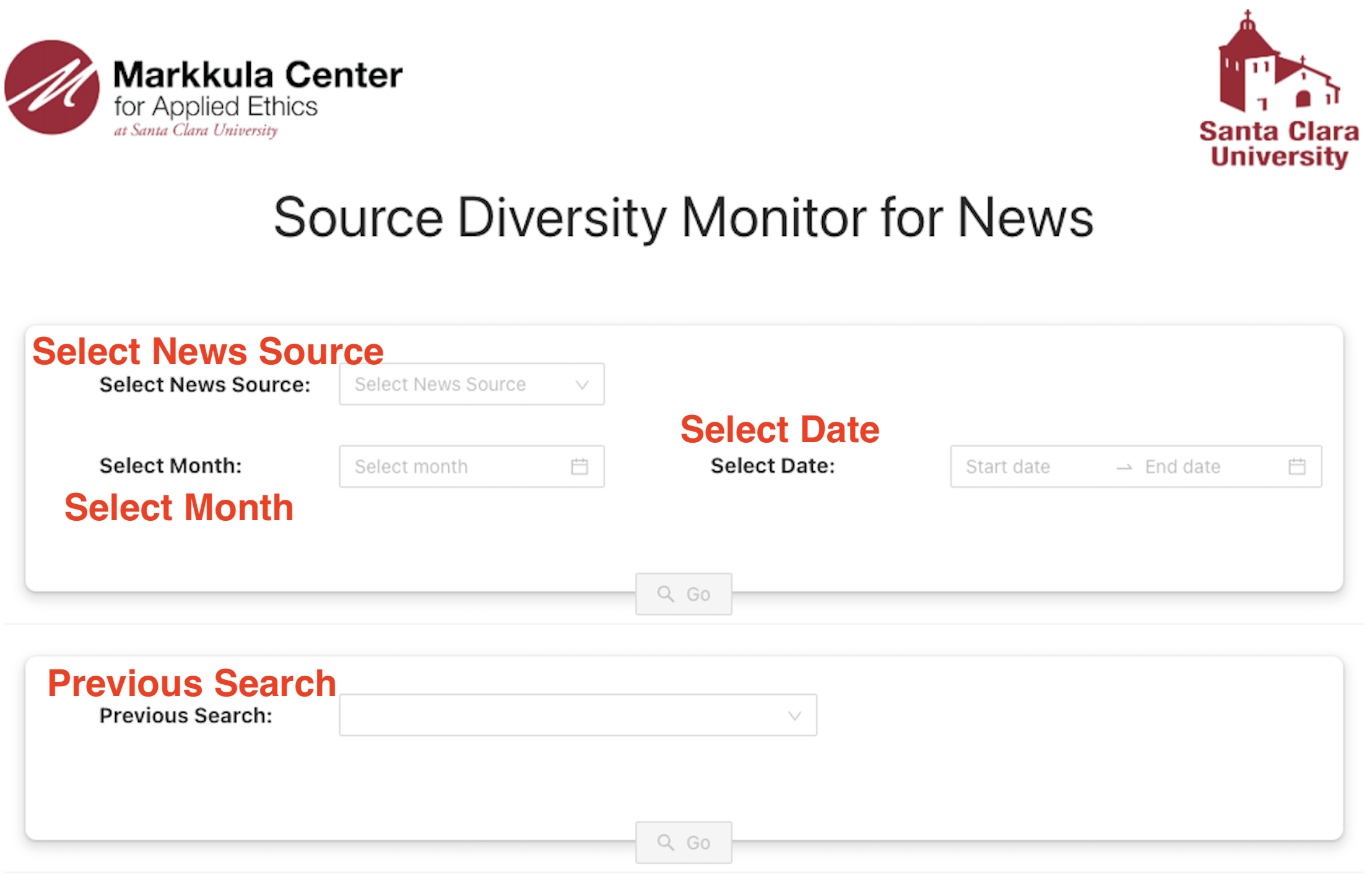}
 \caption{The user interface of the Web Monitor App.}
 \label{fig:selection}
\end{figure}

\subsection{Web Monitor App}
If users want to go beyond a single newsroom, they can use the web monitor that DIANES offers to investigate the DEI data across multiple sites as shown in Figure \ref{fig:selection}. At full scale, the monitor system will be able to host source-diversity data for up to 7,000 U.S. sites. The resulting dashboard is similar to the one illustrated in Figure \ref{fig:monthproportion} with some extra information such as the statistics about the top quoted organizations.

%\begin{figure}[ht]
%\centering
% \includegraphics[width=\linewidth]{monitor.png}
% \caption{The Web Monitor App displays monthly DEI annotated data for all the articles from a news site for selected month.}
% \label{fig:monitor}
%\end{figure}

\subsection{Annotation API}

With the annotation API, users can send a request for any single article, and then receive the structured information about the quotes, speakers, titles, organizations, and community representations. The results are in JSON and can help DEI application developers access the relevant DEI information extracted from any given article to develop their own downstream DEI services and interfaces.

\section{Conclusion and future work}

DIANES is the first DEI audit toolkit that can analyze news articles and visualize source-diversity proportions for quotes on demand, by leveraging the recent advances of natural language processing and information retrieval. It provides easy-to-use user interfaces for reporters, editors, and DEI practitioners to monitor and track the diversity of the news sourcing. 

DIANES is currently tested in the field by several newsrooms. We are improving the toolkit based on the feedback from the users. For example, we will add a feature to allow reporters to override the information extracted and predicted by DIANES when they see fit. This manual correction will not only ensure the high quality of the DEI audit but also provide valuable training data for machine learning models to further improve the performance. In addition, the results of race detection on some minority groups were noticeably worse than those on the majority group due to the limited training data available for the minority races. We will address the imbalanced classification problem by exploring new loss functions and utilizing data augmentation techniques.

\begin{acks}
This project is supported by Google News Initiative and Facebook Research. Prior funding from News Quality Initiative was used to build the custom CoreNLP-based kernel routines to process news writing and build quote annotations data, which was done by Louise Li and Xuyang Wu at Santa Clara University. 
We also knowledge news dataset inputs and review from Laura Moorhead, Associate Professor or Journalism, San Francisco State University.  
\end{acks}

\balance

\bibliographystyle{ACM-Reference-Format}
\bibliography{main}


\begin{thebibliography}{16}

%%% the Web address.

\ifx \showCODEN    \undefined \def \showCODEN     #1{\unskip}     \fi
\ifx \showDOI      \undefined \def \showDOI       #1{#1}\fi
\ifx \showISBNx    \undefined \def \showISBNx     #1{\unskip}     \fi
\ifx \showISBNxiii \undefined \def \showISBNxiii  #1{\unskip}     \fi
\ifx \showISSN     \undefined \def \showISSN      #1{\unskip}     \fi
\ifx \showLCCN     \undefined \def \showLCCN      #1{\unskip}     \fi
\ifx \shownote     \undefined \def \shownote      #1{#1}          \fi
\ifx \showarticletitle \undefined \def \showarticletitle #1{#1}   \fi
\ifx \showURL      \undefined \def \showURL       {\relax}        \fi
% invisible to TeX
\providecommand\bibfield[2]{#2}
\providecommand\bibinfo[2]{#2}
\providecommand\natexlab[1]{#1}
\providecommand\showeprint[2][]{arXiv:#2}

\bibitem[\protect\citeauthoryear{Angeli, Zhong, Chen, Chaganty, Bolton,
  Johnson, Pasupat, Gupta, and Manning}{Angeli et~al\mbox{.}}{2015}]%
        {angeli2015bootstrapped}
\bibfield{author}{\bibinfo{person}{Gabor Angeli}, \bibinfo{person}{Victor
  Zhong}, \bibinfo{person}{Danqi Chen}, \bibinfo{person}{Arun~Tejasvi
  Chaganty}, \bibinfo{person}{Jason Bolton}, \bibinfo{person}{Melvin Johnson},
  \bibinfo{person}{Panupong Pasupat}, \bibinfo{person}{S. Gupta}, {and}
  \bibinfo{person}{Christopher~D. Manning}.} \bibinfo{year}{2015}\natexlab{}.
\newblock \showarticletitle{Bootstrapped Self Training for Knowledge Base
  Population}.
\newblock \bibinfo{journal}{\emph{Theory and Applications of Categories}}.
\newblock


\bibitem[\protect\citeauthoryear{Artwick}{Artwick}{2014}]%
        {artwick2014news}
\bibfield{author}{\bibinfo{person}{Claudette~G Artwick}.}
  \bibinfo{year}{2014}\natexlab{}.
\newblock \showarticletitle{News sourcing and gender on Twitter}.
\newblock \bibinfo{journal}{\emph{Journalism}} \bibinfo{volume}{15},
  \bibinfo{number}{8} (\bibinfo{year}{2014}), \bibinfo{pages}{1111--1127}.
\newblock


\bibitem[\protect\citeauthoryear{Asr, Mazraeh, Lopes, Gautam, Gonzales, Rao,
  and Taboada}{Asr et~al\mbox{.}}{2021}]%
        {asr2021gender}
\bibfield{author}{\bibinfo{person}{Fatemeh~Torabi Asr},
  \bibinfo{person}{Mohammad Mazraeh}, \bibinfo{person}{Alexandre Lopes},
  \bibinfo{person}{Vasundhara Gautam}, \bibinfo{person}{Junette Gonzales},
  \bibinfo{person}{Prashanth Rao}, {and} \bibinfo{person}{Maite Taboada}.}
  \bibinfo{year}{2021}\natexlab{}.
\newblock \showarticletitle{The gender gap tracker: Using natural language
  processing to measure gender bias in media}.
\newblock \bibinfo{journal}{\emph{PloS one}} \bibinfo{volume}{16},
  \bibinfo{number}{1} (\bibinfo{year}{2021}).
\newblock


\bibitem[\protect\citeauthoryear{Brown, Bybee, Wearden, and Straughan}{Brown
  et~al\mbox{.}}{1987}]%
        {brown1987invisible}
\bibfield{author}{\bibinfo{person}{Jane~Delano Brown}, \bibinfo{person}{Carl~R
  Bybee}, \bibinfo{person}{Stanley~T Wearden}, {and}
  \bibinfo{person}{Dulcie~Murdock Straughan}.} \bibinfo{year}{1987}\natexlab{}.
\newblock \showarticletitle{Invisible power: Newspaper news sources and the
  limits of diversity}.
\newblock \bibinfo{journal}{\emph{Journalism Quarterly}} \bibinfo{volume}{64},
  \bibinfo{number}{1} (\bibinfo{year}{1987}), \bibinfo{pages}{45--54}.
\newblock


\bibitem[\protect\citeauthoryear{Das and Paik}{Das and Paik}{2021}]%
        {das2021context}
\bibfield{author}{\bibinfo{person}{Sudeshna Das} {and} \bibinfo{person}{Jiaul~H
  Paik}.} \bibinfo{year}{2021}\natexlab{}.
\newblock \showarticletitle{Context-sensitive gender inference of named
  entities in text}.
\newblock \bibinfo{journal}{\emph{Information Processing \& Management}}
  \bibinfo{volume}{58}, \bibinfo{number}{1} (\bibinfo{year}{2021}).
\newblock


\bibitem[\protect\citeauthoryear{Fu}{Fu}{2021}]%
        {DEX2021}
\bibfield{author}{\bibinfo{person}{Angela Fu}.}
  \bibinfo{year}{2021}\natexlab{}.
\newblock \bibinfo{booktitle}{\emph{New tool allows NPR to track source
  diversity in real time}}.
\newblock
\urldef\tempurl%
\url{https://www.poynter.org/reporting-editing/2021/new-tool-allows-npr-to-track-source-diversity-in-real-time/}
\showURL{%
\tempurl}


\bibitem[\protect\citeauthoryear{Macharia, O'Connor, and Ndangam}{Macharia
  et~al\mbox{.}}{2010}]%
        {macharia2010makes}
\bibfield{author}{\bibinfo{person}{Sarah Macharia}, \bibinfo{person}{Dermot
  O'Connor}, {and} \bibinfo{person}{Lilian Ndangam}.}
  \bibinfo{year}{2010}\natexlab{}.
\newblock \bibinfo{booktitle}{\emph{Who makes the news?: Global media
  monitoring project 2010}}.
\newblock \bibinfo{publisher}{World Association for Christian Communication
  Toronto, Canada}.
\newblock


\bibitem[\protect\citeauthoryear{Manning, Surdeanu, Bauer, Finkel, Bethard, and
  McClosky}{Manning et~al\mbox{.}}{2014}]%
        {manning-etal-2014-stanford}
\bibfield{author}{\bibinfo{person}{Christopher~D Manning},
  \bibinfo{person}{Mihai Surdeanu}, \bibinfo{person}{John Bauer},
  \bibinfo{person}{Jenny Finkel}, \bibinfo{person}{Steven Bethard}, {and}
  \bibinfo{person}{David McClosky}.} \bibinfo{year}{2014}\natexlab{}.
\newblock \showarticletitle{The {S}tanford {C}ore{NLP} Natural Language
  Processing Toolkit}. In \bibinfo{booktitle}{\emph{Proceedings of the 52nd
  Annual Meeting of the Association for Computational Linguistics: System
  Demonstrations}}. \bibinfo{publisher}{Association for Computational
  Linguistics}, \bibinfo{address}{Baltimore, Maryland},
  \bibinfo{pages}{55--60}.
\newblock


\bibitem[\protect\citeauthoryear{Muzny, Fang, Chang, and Jurafsky}{Muzny
  et~al\mbox{.}}{2017}]%
        {muzny2017two}
\bibfield{author}{\bibinfo{person}{Grace Muzny}, \bibinfo{person}{Michael
  Fang}, \bibinfo{person}{Angel Chang}, {and} \bibinfo{person}{Dan Jurafsky}.}
  \bibinfo{year}{2017}\natexlab{}.
\newblock \showarticletitle{A two-stage sieve approach for quote attribution}.
  In \bibinfo{booktitle}{\emph{Proceedings of the 15th Conference of the
  European Chapter of the Association for Computational Linguistics}}.
  \bibinfo{pages}{460--470}.
\newblock


\bibitem[\protect\citeauthoryear{Nivre, De~Marneffe, Ginter, Goldberg, Hajic,
  Manning, McDonald, Petrov, Pyysalo, Silveira, et~al\mbox{.}}{Nivre
  et~al\mbox{.}}{2016}]%
        {nivre2016universal}
\bibfield{author}{\bibinfo{person}{Joakim Nivre},
  \bibinfo{person}{Marie-Catherine De~Marneffe}, \bibinfo{person}{Filip
  Ginter}, \bibinfo{person}{Yoav Goldberg}, \bibinfo{person}{Jan Hajic},
  \bibinfo{person}{Christopher~D Manning}, \bibinfo{person}{Ryan McDonald},
  \bibinfo{person}{Slav Petrov}, \bibinfo{person}{Sampo Pyysalo},
  \bibinfo{person}{Natalia Silveira}, {et~al\mbox{.}}}
  \bibinfo{year}{2016}\natexlab{}.
\newblock \showarticletitle{Universal dependencies v1: A multilingual treebank
  collection}. In \bibinfo{booktitle}{\emph{Proceedings of the Tenth
  International Conference on Language Resources and Evaluation (LREC)}}.
  \bibinfo{pages}{1659--1666}.
\newblock


\bibitem[\protect\citeauthoryear{Reich}{Reich}{2010}]%
        {reich2010source}
\bibfield{author}{\bibinfo{person}{Zvi Reich}.}
  \bibinfo{year}{2010}\natexlab{}.
\newblock \showarticletitle{Source credibility as a journalistic work tool}.
\newblock In \bibinfo{booktitle}{\emph{Journalists, sources, and credibility}}.
  \bibinfo{publisher}{Routledge}, \bibinfo{pages}{31--48}.
\newblock


\bibitem[\protect\citeauthoryear{Santamar{\'\i}a and
  Mihaljevi{\'c}}{Santamar{\'\i}a and Mihaljevi{\'c}}{2018}]%
        {santamaria2018comparison}
\bibfield{author}{\bibinfo{person}{Luc{\'\i}a Santamar{\'\i}a} {and}
  \bibinfo{person}{Helena Mihaljevi{\'c}}.} \bibinfo{year}{2018}\natexlab{}.
\newblock \showarticletitle{Comparison and benchmark of name-to-gender
  inference services}.
\newblock \bibinfo{journal}{\emph{PeerJ Computer Science}}  \bibinfo{volume}{4}
  (\bibinfo{year}{2018}).
\newblock


\bibitem[\protect\citeauthoryear{Schuster and Paliwal}{Schuster and
  Paliwal}{1997}]%
        {schuster1997bidirectional}
\bibfield{author}{\bibinfo{person}{Mike Schuster} {and}
  \bibinfo{person}{Kuldip~K Paliwal}.} \bibinfo{year}{1997}\natexlab{}.
\newblock \showarticletitle{Bidirectional recurrent neural networks}.
\newblock \bibinfo{journal}{\emph{IEEE transactions on Signal Processing}}
  \bibinfo{volume}{45}, \bibinfo{number}{11} (\bibinfo{year}{1997}),
  \bibinfo{pages}{2673--2681}.
\newblock


\bibitem[\protect\citeauthoryear{Schuster and Manning}{Schuster and
  Manning}{2016}]%
        {schuster2016enhanced}
\bibfield{author}{\bibinfo{person}{Sebastian Schuster} {and}
  \bibinfo{person}{Christopher~D Manning}.} \bibinfo{year}{2016}\natexlab{}.
\newblock \showarticletitle{Enhanced english universal dependencies: An
  improved representation for natural language understanding tasks}. In
  \bibinfo{booktitle}{\emph{Proceedings of the Tenth International Conference
  on Language Resources and Evaluation (LREC)}}. \bibinfo{pages}{2371--2378}.
\newblock


\bibitem[\protect\citeauthoryear{Sood and Laohaprapanon}{Sood and
  Laohaprapanon}{2018}]%
        {sood2018predicting}
\bibfield{author}{\bibinfo{person}{Gaurav Sood} {and} \bibinfo{person}{Suriyan
  Laohaprapanon}.} \bibinfo{year}{2018}\natexlab{}.
\newblock \showarticletitle{Predicting race and ethnicity from the sequence of
  characters in a name}.
\newblock \bibinfo{journal}{\emph{arXiv preprint arXiv:1805.02109}}
  (\bibinfo{year}{2018}).
\newblock


\bibitem[\protect\citeauthoryear{Zhang, Chaganty, Paranjape, Chen, Bolton, Qi,
  and Manning}{Zhang et~al\mbox{.}}{2016}]%
        {zhang2016stanford}
\bibfield{author}{\bibinfo{person}{Yuhao Zhang}, \bibinfo{person}{Arun~Tejasvi
  Chaganty}, \bibinfo{person}{Ashwin Paranjape}, \bibinfo{person}{Danqi Chen},
  \bibinfo{person}{Jason Bolton}, \bibinfo{person}{Peng Qi}, {and}
  \bibinfo{person}{Christopher~D Manning}.} \bibinfo{year}{2016}\natexlab{}.
\newblock \showarticletitle{Stanford at TAC KBP 2016: Sealing pipeline leaks
  and understanding chinese}. In \bibinfo{booktitle}{\emph{TAC}}.
\newblock


\end{thebibliography}

\end{document}